\documentstyle[12pt]{article}
\begin{document}
\baselineskip 8mm
\title{Sliding on the inside of a Conical Surface}
\author{
R. L\'{o}pez-Ruiz$^{\ddag}$ and A. F. Pacheco$^{\dag}$,\\
 \small $^{\ddag}$ DIIS - \'Area de Ciencias de la Computaci\'on, \\
 \small $^{\dag}$ Departamento de F\'{\i}sica Te\'orica, \\
 \small Faculty of Sciences - University of Zaragoza, \\
 \small 50009 - Zaragoza (Spain).
 }
 \date{ }

\maketitle
\begin{center} {\bf Abstract} \end{center}
We analyze the frictionless motion of a point-like particle that
slides under gravity on an inverted conical surface. This motion 
is studied for arbitrary initial
conditions and a general relation, valid within $13\%$,
between the periods of radial and angular oscillations, that holds
for both small and high energy trajectories, is obtained. 
This relation allows us
to identify the closed orbits of the system. The virtues of this
model to illustrate pedagogically, how in a physical system, the
energy is transferred between different modes, are also emphasized.
Two easy identification criteria for this type of motion with
potential interest in industrial design are obtained.

{\small Electronic mail: $^{\ddag}$ rilopez@posta.unizar.es ;
$^{\dag}$ amalio@posta.unizar.es}

\newpage

\section{Introduction}
\label{sec:intro}

The conical shape is ubiquitous in nature. Many important large-scale 
natural formations such as the profile of volcanoes 
\cite{turcotte,hancock} or the 
borders  of tornadoes present this universal form \cite{bluestein}. 
On a smaller scale, different devices such as silos, loudspeakers, 
funnels, pipes, particle  precipitators or centrifugal separators
have been designed  by man using this geometry to perform diverse
technical operations \cite{lalor,masters,xiang,alhawaj}.
Don't let us forget other basic devices with conical design 
such as wash-hand basins and lavatories. 
Vortices in fluids appear in some cases
as evolving cone-like structures \cite{fluids}, and recently the 
avalanches of two-dimensional automata modeling the surface 
of a sandpile have received considerable attention 
from the scientific community \cite{bak}. 
Although the physical mechanisms
behind all these complex phenomena are different, the 
coincidence and the ubiquity of the conical form 
in all of them is remarkable.

Bodies falling over inclined surfaces under the influence 
of gravity are also encountered in every day life. Think, for instance,
of a tobbogan, a roller coaster or a snow-board car. These systems are 
modeled  in general physics courses as
point-like particles sliding over those surfaces under the action 
of gravitational and surface reaction forces. 

In this work, we consider the mechanical problem of a point-like 
particle of mass $m$ sliding on the inside of a smooth cone of 
semi-vertical angle $\phi_0$, whose axis points vertically upward 
(see Fig. 1), as a first step to the understanding of those more 
complicated hydrodynamical and granular systems spiraling on the 
inside of a conic surface.
This problem is proposed by T.W.B. Kibble in his
well known Classical Mechanics textbook \cite{kibble}. There, the
student is asked: (1) to find the Hamiltonian function, using
the distance $r$ from the vertex and the azimuth angle $\theta$ as
generalized coordinates ; (2) to show that circular motion is
possible for any value of $r$, and to determine the corresponding
angular velocity $\omega_0$ ; (3) to deduce the angle
$\bar{\phi}_0$ where the frequency of the small radial
oscillations about the circular motion is also $\omega_0$.

Setting the coordinate frame at the cone vertex, denoting the
vertical direction as the $OZ$ axis and using $r$ and $\theta$ as
the generalized coordinates, we deduce that the Hamiltonian function 
for this problem is
\begin{equation}
{\cal H} = \frac{p_r^2}{2m} +
\frac{p_{\theta}^2}{2mr^2\sin^2\phi_0} + mgr\cos\phi_0,
\label{eq:H}
\end{equation}
where $p_r$ and $p_{\theta}$ stand for the canonically conjugated
momenta of $r$ and $\phi_0$ respectively. Besides, we have assumed
that the level of reference for the potential energy is $z=0$, and
the constant value of gravity is denoted by $g$.\newline As the
coordinate $\theta$ does not explicitly appear in $\cal H$,
$\theta$ is a {\it hidden} variable and from the equations of
motion we deduce that $p_{\theta}$ is a constant of the motion.
Physically this is clear because there are only two forces acting
on the particle, one being gravity and the other the force of
reaction exerted by the surface which is perpendicular to the
surface. Thus, both forces are located in the vertical plane
formed by the $OZ$ axis and the position vector of the particle,
and therefore the momentum of these forces is always horizontal and
hence the vertical component of the angular momentum is conserved.
In other words $p_{\theta}$ is a constant.\newline In consequence,
we define the effective radial potential energy $U(r)$ for this
problem as
\begin{equation}
U(r) = \frac{L_z^2}{2mr^2\sin^2{\phi_0}} + mgr\cos\phi_0,
\label{eq:U_r}
\end{equation}
where $L_z$ is the constant value assumed for $p_{\theta}$.
\newline $U(r)$ for $L_z=1$ and $\phi_0={\pi\over 4}$, is plotted in Fig.
2. Thus $U(r)$ is a confining potential and the particle, no
matter what its energy is, cannot escape from the cone. This confinement
is due to the effect of gravity for large $r$ and to the reaction
of the conical surface as a consequence of the acquired rotational
kinetic energy for small $r$. In both cases, the conservation of
$p_{\theta}$ limits the transference between gravitational energy
and kinetic energy as will be discussed in detail below. From Eq.
(\ref{eq:U_r}) we deduce the radial position $r_0$ where $U$ has
its minimum $U_0$,
\begin{equation}
r_0 =
\left[\frac{L_z^2}{m^2g\sin^2\phi_0\cos\phi_0}\right]^{1\over 3}.
\label{eq:r_0}
\end{equation}
Thus, for any pair $\phi_0$ and $L_z$, there exists a unique
circular motion at the distance $r=r_0$, with angular velocity
\begin{equation}
\omega_0 = \frac{L_z}{mr_0^2\sin^2\phi_0}. \label{eq:w_0}
\end{equation}
Now, imparting a small $L_z$-conserving energy perturbation
$\Delta$ to this circular motion, up to an energy $E=U_0+\Delta$,
one generates an orbit with small radial oscillation around $r_0$
(see Fig. 2). The frequency $\omega_r$ of this oscillation is
\begin{equation}
w_r = \sqrt{\frac{1}{m}\left[\frac{d^2U}{dr^2}\right]_{r=r_0}} =
\sqrt{\frac{1}{m}\;\frac{3L_z^2}{mr_0^4\sin^2\phi_0}}.
\label{eq:w_r}
\end{equation}
Thus the radial time period $T_r=\frac{2\pi}{\omega_r}$ coincides
with the period of the circular motion when
$T_r=\frac{2\pi}{\omega_0}$. This condition leads to
\begin{equation}
\bar{\phi}_0 = \sin^{-1}(1/\sqrt{3}) \simeq 35.3^0,
\label{eq:phi0}
\end{equation}
which completes the answer to the three academic 
questions proposed at the
beginning of this Section. 

In Section \ref{sec:oscillations}, we will study the
equations of motion and the trajectories of this system with
arbitrary energies and will obtain a general relation between the
periods of radial and angular oscillations. 
In Section \ref{sec:universal},
several universal geometrical and dynamical relations for a general
trajectory are derived. The rich energy transfer process acting 
in this system is also studied. 
In Section \ref{sec:conclusions}, we state the conclusions.
Finally, in the Appendix \ref{appendix1}, we will study the necessary
conditions for having periodicity in a general orbit.

\section{Small and Large Radial Oscillations}
\label{sec:oscillations}

\subsection{Equations of the system}
As said in Section \ref{sec:intro}, the dynamics of the particle
inside the cone can be expressed in the generalized coordinates
$(r,\theta)$ (see Fig. 1). If $v$ is the velocity of the particle,
its kinetic energy $\cal T$ is written in these coordinates as
\begin{equation}
{\cal T} = {1\over 2}\; mv^2 = {1\over 2}\; m(\dot{r}^2 +
r^2\dot{\theta}^2\sin^2\phi_0), \label{eq:T}
\end{equation}
where the dot means  the time derivative. The potential energy
$\cal V$ of the particle in the gravitational field is
\begin{equation}
{\cal V} = mgz = mgr\cos\phi_0.
 \label{eq:V}
\end{equation}
Then the Lagrangian function ${\cal L}$ for this system in these
particular coordinates appears as
\begin{equation}
{\cal L} = {\cal T} - {\cal V} = {1\over 2}\; m(\dot{r}^2 +
r^2\dot{\theta}^2\sin^2\phi_0) - mgr\cos\phi_0. \label{eq:L}
\end{equation}
The generalized momenta are
\begin{eqnarray}
p_r = \frac{\partial\cal L}{\partial\dot{r}} & = & m\dot{r},
\label{eq:momenta1}\\ p_{\theta} = \frac{\partial\cal
L}{\partial\dot{\theta}} & = & mr^2\dot{\theta}\sin^2\phi_0.
\label{eq:momenta2}
\end{eqnarray}
If we substitute these new coordinates $(p_r,p_{\theta})$ in the
function
\begin{displaymath}
{\cal H}\; = \;\dot{r}p_r + \dot{\theta}p_{\theta} - {\cal L}
\;=\; {\cal T} + {\cal V},
\end{displaymath}
the Hamiltonian $\cal H$ shown in Eq. (\ref{eq:H}) is obtained.

The equations of motion of the particle are determined by the
Lagrange's equations:
\begin{eqnarray*}
\frac{d}{dt}\left(\frac{\partial\cal L}{\partial\dot{r}}\right) -
\frac{\partial\cal L}{\partial r} & = & 0, \\
\frac{d}{dt}\left(\frac{\partial\cal
L}{\partial\dot{\theta}}\right) - \frac{\partial\cal
L}{\partial\theta} & = & 0. \label{eq:Lagrange}
\end{eqnarray*}
The first of these equations gives us the evolution of the
particle in the radial direction. It yields
\begin{equation}
\ddot{r} - r\dot{\theta}^2\sin^2\phi_0 + g\cos\phi_0 = 0.
\label{eq:radial1}
\end{equation}
And the second puts in evidence an invariant of motion because
$\cal L$ is independent on the angular variable $\theta$:
\begin{equation}
mr^2\dot{\theta}\sin^2\phi_0 = cte. = L_z.
\label{eq:angular}
\end{equation}
This dynamical constant, $L_z$, is the vertical component of the
angular momentum.\newline If we substitute the value of
$\dot{\theta}=L_z/(mr^2\sin^2\phi_0)$ in Eq. (\ref{eq:radial1}),
the radial evolution is uncoupled from its angular dependence
which only remains present through the constant $L_z$,
\begin{equation}
\ddot{r} - \left(\frac{L_z}{m\sin\phi_0}\right)^2\frac{1}{r^3} +
g\cos\phi_0 = 0.
 \label{eq:radial}
\end{equation}
This last equation corresponds to an integrable nonlinear
oscillator in the radial direction. After integrating this motion
the angular part of the dynamics is obtained through Eq.
(\ref{eq:angular}). Next we analyze the behavior of the radial
oscillator: first for the small oscillations and then for
arbitrary energy perturbations.

\subsection{The radial motion}
We can see at a glance the whole picture of the possible
motions in the radial coordinate if the totality of its integral
curves on the plane $(r,\dot{r})$ are known \cite{andronov}.
\newline Equation (\ref{eq:radial}) can be easily integrated since
the radial and time variables are separated. If we define the
parameters
\begin{eqnarray*}
A & = & \left(\frac{L_z}{m\sin\phi_0}\right)^2, \\
 B & = & g\cos\phi_0,
\end{eqnarray*}
and $h$ is the constant of the first integration of Eq.
(\ref{eq:radial}), we obtain
\begin{equation}
\frac{\dot{r}^2}{2} + V(r) = h, \label{eq:radial-h}
\end{equation}
with
\begin{eqnarray}
V(r) & = & \frac{A}{2r^2} + Br = \frac{U(r)}{m}, \label{eq:V_r} \\
h & = & \frac{\cal H}{m},
\end{eqnarray}
where $U(r)$ is the effective radial potential energy given by Eq.
(\ref{eq:U_r}) and $\cal H$ is the total energy of the system (Eq.
(\ref{eq:H})).

{\bf Circular motion}: Depending on the initial conditions
$(r,\dot{r})_{t=0}$, the energy constant $h$ takes different
values  and a different equi-energy curve is drawn by the system
for every $h$ (see Fig. 3). The value $h=h_0$ for which the
integral curve degenerates in an isolated point, occurs for the
only minimum of the potential $V(r)$, given by the relation
\begin{displaymath}
\frac{dV(r)}{dr} = 0 \;\Rightarrow\; r = r_0 =
\sqrt[3]{\frac{A}{B}},
\end{displaymath}
where $r_0$ was written in Eq. (\ref{eq:r_0}) and represents the
only existing circular orbit of the system for the actual values
of $L_z$ and $\phi_0$. By substituting
$L_z=mr_0^{2}\omega_0\sin^2\phi_0$ in the expression
(\ref{eq:r_0}), the angular frequency $\omega_0$ of the circular
motion given by Eq. (\ref{eq:w_0}) is found to be
\begin{displaymath}
\omega_0^2 = \frac{g\cos\phi_0}{r_0\sin^2\phi_0}.
\end{displaymath}
It is curious that the angle $\phi_0$ which minimizes $r_0$ for a
$L_z$ fixed is the complementary angle of $\bar{\phi}_0$, written
in (\ref{eq:phi0}), that is, $\phi_0=54.7^0$. \newline
 {\bf General radial orbit}: For $h<h_0$ motion is not possible and for
$h>h_0$ the phase plane organizes itself as a pattern of closed curves
nested around the equilibrium point $r=r_0$ (Fig. 3). This is a
consequence of the confining property of $V(r)$, which means
\begin{eqnarray*}
\lim_{r\downarrow 0} V(r) \sim \frac{A}{2r^2}
\;\;\stackrel{r\rightarrow 0 }{\longrightarrow} +\infty,\\
 \lim_{r\uparrow \infty} V(r) \sim Br
 \;\;\stackrel{r\rightarrow \infty }{\longrightarrow} +\infty.
\end{eqnarray*}
Hence, given the initial conditions of the system, the particle
inside the cone is oscillating in the radial coordinate between
the minimum, $r_{min}$, and the maximum, $r_{max}$, values of its
radial trajectory (Fig. 2). Taking into account that $r_0^3=A/B$
and that $V(r)$ reaches the same value at these extreme points,
$V(r_{min})=V(r_{max})$, we obtain a global relation of a
particular trajectory:
\begin{equation}
\frac{2\,r_{min}^2r_{max}^2}{r_{min}+r_{max}} = r_0^3.
\label{eq:traject-rel}
\end{equation}

\subsection{Small oscillations} For a value $h_1$ slightly bigger
than $h_0$, the motion of the system is represented approximately
on the phase plane $(r,\dot{r})$ by an ellipse around the singular
point $r=r_0$. In this case, $r_{min}$ and $r_{max}$ are of the
same order of magnitude as $r_0$ and the relation
(\ref{eq:traject-rel}) becomes simplified to
\begin{equation}
r_{min} r_{max} \simeq r_0^2. \label{traject-small}
\end{equation}
If we put $r=r_0+\rho$, with $\rho$ small, and we linearize the
expression (\ref{eq:radial-h}), the equation of a harmonic
oscillator is derived:
\begin{equation}
\ddot{\rho} + \omega_r^2 \rho \simeq 0 \label{eq:harmonic},
\end{equation}
where
\begin{displaymath}
\omega_r^2 = \left[\frac{d^2V}{dr}\right]_{r=r_0} =
\frac{3g\cos\phi_0}{r_0},
\end{displaymath}
as was advanced in Eq. (\ref{eq:w_r}). This approximation is
valid, provided the energy of the perturbation $\Delta=m(h_1-h_0)$
is mainly stored in the radial harmonic oscillation. This is
verified when the condition $\frac{\Delta}{m\omega_r^2\rho^2}\sim
1$ is fulfilled. \newline The relation between this radial
frequency and the frequency of the circular motion is
\begin{equation}
\frac{w_r^2}{w_0^2} = 3 \sin^2{\phi_0}. \label{eq:rel-w}
\end{equation}
If we define the respective time periods, $T_r=\frac{2\pi}{w_r}$
and $T_0=\frac{2\pi}{w_0}$, the latter expression implies that
\begin{displaymath}
\frac{T_0}{T_r} = \sqrt{3}\sin\phi_0,
\end{displaymath}
which indicates that the angular part of the motion is faster than
the radial when $\phi_0<\bar{\phi}_0$. In this range, the particle
will perform many revolutions around the cone axis for each radial
oscillation.

An orbit will represent a periodic motion, and thus, a closed
trajectory in the four-dimensional phase space
$(r,\theta,\dot{r},\dot{\theta})$, when the ratio
$\omega_r/\omega_0$ is a rational number (see Appendix
\ref{appendix1}). In the case of small radial oscillations the
perturbed circular orbit will be periodic if there exists a pair
$(p,q)$ of integers verifying
\begin{equation}
\frac{p^2}{q^2} = 3\sin^2\phi_0. \label{eq:p-q}
\end{equation}
This means that for each pair $(p,q)$, with $p<\sqrt{3}\;q$, there
is an angle $\phi_0$ of the cone opening for which the perturbed
circular orbits are periodic. If this last relation (\ref{eq:p-q})
is not satisfied for any pair $(p,q)$, the dynamics is
quasi-periodic and fill up the torus on which it develops. As an
example, if $p=q=1$, then $\phi_0=\bar{\phi}_0\simeq 35.3^0$ (Eq.
(\ref{eq:phi0})) and the small radial-oscillating trajectories are
ellipses on the cone.

The wave shape of the perturbed circular orbit in the plane
$(\theta, r)$ is not sinusoidal as can be checked in Fig. 4.
The trajectory covers a longer $\theta$-angular distance when its
radial coordinate is under the value $r=r_0$ than when it is in
the upper region of $r_0$. A look at the first order approximation
to the orbit explains this fact. If we write
\begin{displaymath}
r(t)\simeq r_0 + \rho\cos(\omega_r t),
\end{displaymath}
then, from Eq. (\ref{eq:angular}), we obtain
\begin{displaymath}
\theta(t)\simeq \omega_0 t -
2\left(\frac{\omega_0}{\omega_r}\right)
\left(\frac{\rho}{r_0}\right)\sin(\omega_r t).
\end{displaymath}
A whole oscillation in $r$ is made when $\theta$ runs on the
interval $\frac{\omega_0}{\omega_r}\cdot\left[0, 2\pi\right]$. If
the perturbation in the $\theta$-coordinate is not considered, the
orbit $r(\theta)$ is sinusoidal and satisfies $r(\theta)<r_0$ when
$\theta\in\frac{\omega_0}{\omega_r}\cdot\left[\frac{\pi}{2},
\frac{3\pi}{2}\right]$. Actually, a simple calculation shows that
the small perturbation forces the dynamics to stay under $r=r_0$
on an enlarged $\theta$-interval, namely
$\frac{\omega_0}{\omega_r}\cdot\left[\frac{\pi}{2}-\frac{2\rho}{r_0},
\frac{3\pi}{2}+\frac{2\rho}{r_0}\right]$. This shape wave
deformation continues in the same direction by increasing the
energy of the particle, in such a way that the motion verifies
$r(\theta)<r_0$ for the biggest part of the $\theta$-interval
$\frac{\omega_0}{\omega_r}\cdot\left[0, 2\pi\right]$, except for a
very narrow region where $r(\theta)$ completes the oscillation and
reaches $r_{\max}$ with a sharp peak in the $(\theta, r)$
representation.

\subsection{Large oscillations}
\label{sec:large-osc} For $h\gg h_0$, large oscillations in the
radial coordinate are obtained (see Fig. 3). The trajectories
become ovoid-like reaching the maximum radial velocity
$\dot{r}_{max}$ for $r=r_0$: $\dot{r}_{max}=\dot{r}(r_0)$. As
$r_{max}\gg r_{min}$, the relation ($\ref{eq:traject-rel}$)
simplifies to
\begin{displaymath}
2\,r_{min}^2r_{max} \simeq r_0^3.
\end{displaymath}
Finding the relation between the radial and angular frequencies
requires a more elaborated calculation in this case. We proceed to
integrate Eqs. (\ref{eq:radial}) and (\ref{eq:angular}) when
$h\rightarrow\infty$, in order to obtain the asymptotic value of
$\omega_r/\omega_{\theta}$.

We consider a half-oscillation between the extreme points,
$r_{min}$ and $r_{max}$, of an orbit. We divide this interval into
two parts: $(r_{min},r_0)$ and $(r_0,r_{max})$, which are covered
in the time intervals: $(0,t_1)$ and $(t_1,t_2)$, respectively. As
the variables in the radial equation (\ref{eq:radial}) are
separated, we write
\begin{equation}
\int_{0}^{t_2} dt = \int_{r_{min}}^{r_{max}}
\frac{dr}{\sqrt{2(h-V(r))}} = \int_{r_{min}}^{r_{max}} \frac{ r
dr}{\sqrt{2hr^2-2Br^3-A}}. \label{eq:integ-radial}
\end{equation}
This elliptic integral cannot be put in elemental functions. When
$h\rightarrow\infty$, the potential $V(r)$ behaves as
\begin{eqnarray*}
V(r)\simeq V_1(r) & = & \frac{A}{2r^2} \;\;\;\;\hbox{for}\;\;
r_{min}<r<r_0,
\\ V(r)\simeq V_2(r) & = & Br \;\;\;\;\hbox{for}\;\;\;
r_0<r<r_{max},
\end{eqnarray*}
and, $r_{min}$ and $r_{max}$ are now easily calculated,
\begin{eqnarray*}
V_1(r_{min})\simeq h & \Rightarrow & r_{min}\sim
\sqrt{\frac{A}{2h}},
\\ V_2(r_{max})\simeq h & \Rightarrow & r_{max}\sim \frac{h}{B}.
\end{eqnarray*}
Under these assumptions an approximated value of integral
(\ref{eq:integ-radial}) is obtained. The error is basically
concentrated in the region $(r_0-\Delta r, r_0+\Delta r)$.
However, because velocity is maximal in $r_0$ and tends to
infinity when $h\rightarrow\infty$, the time invested by the
system in that region tends to zero and thus the error vanishes.

Substituting $V(r)$ by $V_1(r)$ and integrating expression
(\ref{eq:integ-radial}) in the interval $(r_{min},r_0)$, we get
\begin{equation}
t_1\simeq
\frac{1}{\sqrt{2h}}\sqrt{r_0^2-r_{min}^2}\;\;\;\stackrel{h\rightarrow\infty}
{\longrightarrow}\;\;\; \frac{r_0}{\sqrt{2h}}. \label{eq:t1}
\end{equation}
Making $V(r)\simeq V_2(r)$ in the region $(r_0, r_{max})$, the
integral (\ref{eq:integ-radial}) is approximated by
\begin{equation}
t_2-t_1\simeq
\sqrt{\frac{2}{B}}\sqrt{r_{max}-r_0}\;\;\;\stackrel{h\rightarrow\infty}
{\longrightarrow}\;\;\; \frac{\sqrt{2h}}{B}. \label{eq:t2}
\end{equation}
Thus the system spends most of the oscillation time in the upper
region of the circular orbit $r=r_0$, and it tends to infinity
when $h\rightarrow\infty$.

The angular part of the motion is derived from Equation
(\ref{eq:angular}),
\begin{equation}
\int_0^{\theta_2} d\theta = \frac{L_z}{m\sin^2\phi_0} \int_0^{t_2}
\frac{dt}{r^2(t)}, \label{eq:integ-angular}
\end{equation}
where the integral is performed in the angular regions
$(0,\theta_1)$ and $(\theta_1,\theta_2)$ which are covered when
time runs in the intervals $(0,t_1)$ and $(t_1,t_2)$,
respectively.\newline Taking into account the relation
\begin{displaymath}
r^2(t)\simeq 2ht^2 + r_{min}^2 \;\;\;\;\hbox{for}\;\;\; 0<t<t_1,
\end{displaymath}
and, using the value of $t=t_1$ given by expression (\ref{eq:t1}),
we obtain $\theta_1$ when $h\rightarrow\infty$:
\begin{equation}
\theta_1\;\; = \;\; \frac{\pi}{2\sin\phi_0}. \label{eq:teta1}
\end{equation}
The behavior in the second interval $(\theta_1,\theta_2)$ is
quite different. Starting from the relation
\begin{displaymath}
t-t_1\;\;\simeq\;\;
\sqrt{\frac{2}{B}}\left(\sqrt{r_{max}-r_0}-\sqrt{r_{max}-r}\right)
\;\;\;\hbox{for}\;\;\; t_1<t<t_2,
\end{displaymath}
$r(t)$ can be derived. As $\frac{Bt}{\sqrt{2h}}<1$, expression
(\ref{eq:integ-angular}) is approximately integrated in the
interval $(t_1,t_2)$. We get in the limit $h\rightarrow\infty$:
\begin{equation}
\theta_2-\theta_1 \;\;\simeq\;\; \frac{1}{r_0}\sqrt{\frac{2}{h}},
\label{eq:teta2}
\end{equation}
which shows that for large energies the dynamics for $r>r_0$ is
projected onto a tight peak in the $(\theta,r)$ plane. Thus most of
the part of the $\theta$-coordinate is covered when the particle
is under the circular orbit $r=r_0$ although the system spends
its time essentially over that circular orbit (see Fig. 5).

Summarizing, if $\frac{T_r}{2}\simeq t_2$ is the semi-period of
the radial oscillation, the frequencies $w_r$ and
$\omega_{\theta}$ for large energies are
\begin{eqnarray*}
\omega_r = \frac{\pi}{T_r/2} & \;\;\simeq\;\; & \frac{2\pi}{t_2},
\\ \omega_{\theta} = \frac{\theta_2}{T_r/2} & \;\;\simeq\;\; &
\frac{\pi/\sin\phi_0}{t_2},
\end{eqnarray*}
which implies that, when $h\rightarrow\infty$, the frequency ratio
is
\begin{equation}
\frac{\omega_r}{\omega_{\theta}} \;\; = \;\; 2\sin\phi_0.
\end{equation}

Comparing this last expression with the frequency ratio
(\ref{eq:rel-w}) for small radial oscillations, a general
behavior of this quantity can be advanced:
\begin{equation}
 \label{eq:univ-rel}
\frac{\omega_r}{\omega_{\theta}} \; = \; k\sin\phi_0
\;\;\;\hbox{with $k$ in the range}\;\; \left\{\begin{array}{c}
 \sqrt{3} < k < 2 \\
 \;\;\updownarrow \mbox{ } \;\;\;\updownarrow \mbox{ } \;\;\updownarrow \\
\;\;\; h_0 < h < \infty  \end{array}\right. .
\end{equation}
This relation will be computationally and analytically discussed
in the next section.

\section{Universal Relations of the Dynamics}
\label{sec:universal}

\subsection{Universal equations}
As has been established in the last section, the evolution of the
particle sliding inside the cone is governed by Eq.
(\ref{eq:radial}) in its radial part and by Eq. (\ref{eq:angular})
in its angular one. These equations are expressed in a form which depends on
the gravitational field and surface geometrical properties and,
evidently, on the initial value of the global dynamical constants.
We proceed now to write them in a universal form by the rescaling of
radial, angular and time variables. In short, all information
on the dynamics will be contained in these reduced equations which
are independent of any 'external' information. To obtain the real
trajectory from these equations it will be enough to undo the
change of variables with the 'external' information: mass,
geometry, field and initial conditions.

If we substitute the value of the vertical angular momentum,
$L_z^2$, given by Eq. (\ref{eq:r_0}) and we perform the change of
variable: $\tilde{r}=r/r_0$, radial and angular equations are
reduced to
\begin{eqnarray}
\ddot{\tilde{r}} + \frac{\omega^2_{r,h_0}}{3} (1 - \tilde{r}^{-3})
& = & 0, \\
 {\dot{\theta}}^2 - \omega_0^2\tilde{r}^{-4} & = & 0,
\end{eqnarray}
where $\omega^2_{r,h_0}$ and $\omega^2_0$ take the values given by
Eqs. (\ref{eq:w_r}) and (\ref{eq:w_0}), respectively. \newline
Rescaling the time variable by
$\tilde{t}=\frac{\omega_{r,h_0}}{\sqrt{3}}t$ and the angular
variable by $\tilde{\theta}=\theta\sin\phi_0$, the {\it
non-dimensional} equations of the motion are found:
\begin{eqnarray}
\ddot{\tilde{r}} + (1 - \tilde{r}^{-3}) & = & 0, \\
 \dot{\tilde{\theta}} - \tilde{r}^{-2} & = & 0.
\end{eqnarray}
Remarkably, in this form, the equations apparently lose any 
characteristic of the system. Every possible trajectory on
the conical surface is projected into a solution of these equations
and, in that sense, we say that they are universal. They contain
all the information about the dynamical behavior of the
particle.\newline In particular, $\tilde{r}= 1$ is the singularity
representing the circular orbit and any other orbit runs between
the extreme values, $\tilde{r}_{min}$ and $\tilde{r}_{max}$, which
satisfy: $0<\tilde{r}_{min}<1$ and $1<\tilde{r}_{max}<\infty$.
Expression (\ref{eq:traject-rel}) is now rewritten as
\begin{equation}
\frac{2\,\tilde{r}_{min}^2\tilde{r}_{max}^2}{\tilde{r}_{min}+\tilde{r}_{max}}
= 1. \label{eq:traject-rel1}
\end{equation}

\subsection{Universal frequency relation}
From the universal equations of the dynamics, we remake now all the
calculations presented in Subsection \ref{sec:large-osc} to show
the universal relation (\ref{eq:univ-rel}). For simplicity in the
notation, we rename the variables: $\tilde{r}\rightarrow r$ and
$\tilde{t}\rightarrow t$.

The energy conservation becomes in the new coordinates
\begin{equation}
\label{eq:radial-e}
\frac{\dot{r}^2}{2} + \tilde{V}(r) = E
\end{equation}
where $\tilde{V}(r) = r+\frac{1}{2r^2}$ and
$E=\frac{h}{gr_0\cos\phi_0}$. Hence, in this representation, the
dynamics settles in the circular orbit when $E=3/2$, and, small or
large oscillations are obtained when the value of the normalized
energy $E$ runs on the interval $(\frac{3}{2},\infty)$.\newline
For an arbitrary energy $E\in(\frac{3}{2},\infty)$, $r_{min}$ is
deduced from the equality:
\begin{displaymath}
\tilde{V}(r_{min}) = E \;\;\rightarrow\;\; r_{min}=r_{min}(E).
\end{displaymath}
The radial coordinate as function of the time, $r(t)$, is
calculated after performing the integral:
\begin{displaymath}
\int_0^t dt = \int^r_{r_{min}} \frac{r
dr}{\sqrt{2Er^2-2r^3-1}}\;\;\rightarrow\;\; r=r(t,E).
\end{displaymath}
For $t=\frac{T_r}{2}$, with $T_r$ the semi-period of the radial
oscillation, $r_{max}$ is reached. Recall that $r_{max}(E)$ is
obtained from Expression (\ref{eq:traject-rel1}). Thus, we get
$T_r(E)$ by solving the equation
\begin{displaymath}
r_{max}(E) = r\left(\frac{T_r}{2},E\right) \;\;\rightarrow\;\;
T_r=T_r(E).
\end{displaymath}
The angle $\tilde{\theta}_2$ covered by the particle during a
radial semi-period is given by
\begin{displaymath}
\tilde{\theta}_2 = \int_0^{\tilde{\theta}_2} d\tilde{\theta} =
\int_0^{T_r/2} \frac{dt}{r^2(t)} \;\;\rightarrow\;\;
\tilde{\theta}_2=\tilde{\theta}_2(E).
\end{displaymath}
Therefore the frequency ratio obeys the following relation:
\begin{equation}
\frac{\omega_r}{\omega_{\theta}} =
\frac{\pi/\frac{T_r}{2}}{\theta_2/\frac{T_r}{2}} =
\frac{\pi}{\tilde{\theta}_2}\;\sin\phi_0 = k(E)\sin\phi_0,
\label{eq:freq-rel}
\end{equation}
where
\begin{equation}
k(E) = \left[\frac{1}{\pi}\int_0^{\frac{T_r(E)}{2}}
\frac{dt}{r^2(t,E)}\right]^{-1}. \label{eq:k(E)}
\end{equation}
As we calculated in the previous section,
\begin{eqnarray*}
k\left(E=\frac{3}{2}\right) & = & \sqrt{3}, \\
 k(E=\infty) & = & 2,
\end{eqnarray*}
and as can be deduced from numerical calculations of $k(E)$
(see Fig. 6), we claim that
\begin{equation}
\sqrt{3}<k(E)<2 \;\;\;\; \hbox{when}\;\;\; \frac{3}{2}<E<\infty,
\end{equation}
which allows us to approximate integral (\ref{eq:k(E)}), valid
within $13\%$, by
\begin{displaymath}
\int_0^{\frac{T_r}{2}} \frac{dt}{r^2(t)}\simeq \frac{\pi}{2},
\end{displaymath}
for any trajectory $r(t)$ of arbitrary energy $E$.

\subsection{Universal dynamical relation}
\label{sec:univ-dyn} We proceed to convert the almost
universal frequency ratio (\ref{eq:univ-rel}) 
in a dynamical relation. If $N$
represents the number of revolutions that the particle performs
around the cone in the half of a radial oscillation, then
$\theta_2=2\pi N$, and expression (\ref{eq:freq-rel}) is now
written as follows:
\begin{equation}
\frac{\omega_r}{\omega_{\theta}} = \frac{\pi/\frac{T_r}{2}}{2\pi N
/\frac{T_r}{2}} = \frac{1}{2 N} = k \sin\phi_0.
\label{eq:freq-rel1}
\end{equation}
Taking the good approximation $k\simeq 2$, we obtain the universal
dynamical relation
\begin{equation}
4N\sin\phi_0 \;\;\simeq\;\; 1.
\end{equation}
In the case of small cone opening, $\phi_0\ll 1$, this simplifies to
\begin{equation}
4 N \phi_0 \;\;\simeq\;\; 1,
\end{equation}
which establishes a simple experimental test to find out if a particle
describing a trajectory within a conical-like surface is truly
governed by the equations under study. That is, if the particle's
dynamics is equivalent to sliding without friction on the surface.

\subsection{Universal geometrical relation}
Following the same line of reasoning as in Section
\ref{sec:univ-dyn} and motivated by the functioning of the
industrial 'cyclone' gas-particle separators \cite{lalor,masters}, 
suppose a motion of
the system developing on a truncated cone such as that drawn in
Fig. 7. The geometrical lengths of the orbit are the maximum and
minimum radii, $R_{min}$ and $R_{max}$, and the vertical height,
$H$, given by
\begin{eqnarray*}
R_{max} & = & r_{max}\sin\phi_0, \\
 R_{min} & = & r_{min}\sin\phi_0 \\
 H & = & \frac{R_{max}-R_{min}}{\tan\phi_0}.
\end{eqnarray*}
By replacing the values of $r_0$ and $L_z$,
\begin{eqnarray*}
L_z & = & m R_{min} v_{down}, \\
 r_0^3 & = &
 \frac{1}{g\cos\phi_0}\left(\frac{R_{min}v_{down}}{\sin\phi_0}\right)^2,
\end{eqnarray*}
in the relation (\ref{eq:traject-rel}), we find a universal
geometrical requirement  for an arbitrary trajectory
\begin{equation}
\frac{R_{max}^2}{R_{min}+R_{max}} =
\frac{v_{down}^2\;\tan\phi_0}{2g}, \label{eq:univ-geom}
\end{equation}
where $v_{down}$ is the linear velocity of the particle at the
bottom of its orbit. \newline Conservation of $L_z$ implies that
\begin{equation}
\frac{R_{max}}{R_{min}} = \frac{v_{down}}{v_{up}},
\end{equation}
with $v_{up}$ the linear velocity at the top of the motion.
Expression (\ref{eq:univ-geom}) can also been written as follows:
\begin{equation}
\frac{R_{min}^2}{R_{min}+R_{max}} =
\frac{v_{up}^2\;\tan\phi_0}{2g}. \label{eq:univ-geom1}
\end{equation}
These universal geometrical relations can be also used to decide if
a physical phenomenon has an underlying dynamics such as
that studied here.

\subsection{Energy transformation}
The kinetic energy ${\cal T}_{\theta}$ of the angular part of the
motion can be written as
\begin{displaymath}
{\cal T}_{\theta} = \frac{1}{2}\;L_z\;\dot{\theta},
\end{displaymath}
and the conservation of energy at a point $P(r,\theta)$ of a
trajectory is the result of the balance between kinetic and
potential energy,
\begin{equation}
-\frac{1}{2}\;m\dot{r}^2 +
\frac{1}{2}\;L_z\;(\dot{\theta}_{down}-\dot{\theta}) = m g \Delta
z,
\end{equation}
where $\dot{r}$ and $\dot{\theta}$ are the radial and angular
velocities at $P$, $\;\;\;\Delta z=(r-r_{min})\cos\phi_0$ is the
vertical distance from $P$ to the bottom of the trajectory and
$\dot{\theta}_{down}$ is the angular velocity at the bottom of the
motion.\newline If we rename $v_r=\dot{r}$ and
$v_{down}=R_{min}\dot{\theta}_{down}$, $\Delta z$ is obtained from
the latter expression,
\begin{equation}
\Delta z \;=\; -\frac{v_r^2}{2g}\; +\;
\frac{v_{down}^2}{2g}\;\left(1-\left(\frac{R_{min}}{R}\right)^2\right),
\end{equation}
where $R=r\sin\phi_0$.

The vertical distance, $H=\Delta z(R_{max})$, covered by the
particle in its motion verifies
\begin{equation}
H =
\frac{v_{down}^2}{2g}\;\left(1-\left(\frac{R_{min}}{R_{max}}\right)^2\right).
\end{equation}
Observe that not all the kinetic energy can be transferred to
potential energy at the top of the orbit. A percentage of the
total energy, given by the factor $\gamma =
\left(\frac{R_{min}}{R_{max}}\right)^2$, remains as angular kinetic
energy in order to conserve $L_z$. This transference is more
efficient when the geometry of the system approaches that of a
plane system because $L_z$ can be conserved by the large radial
component of the trajectory. This happens when $\phi_0$ tends to
$\frac{\pi}{2}$. On the contrary, the transfer of energy is nearly
forbidden and is totally inefficient when the system tends to the
cylindrical form, that is, when $\phi_0$ goes to zero.

The interesting energy-transfer process acting in this system,
between the angular kinetic term, the radial kinetic term and the
potential energy term is not habitual in simple mechanical systems
of only two degrees of freedom. This type of idea typically
appears later when dealing with considerably more complex systems
of three or more degrees of freedom, such as the symmetric top.

Note also that in other problems of a point-like particle sliding
under gravity on surfaces such as planes or cylinders, this rich
energy-transfer does not occur. There the transfer is as in a
bullet motion, i.e., between the 'vertical' kinetic term and the
potential energy term. In the case of sliding on the inside of a
spherical surface the energy transfer process occurs as in the cone
but with two important differences. First, in the sphere, no
matter the value of the energy, the spatial range of the orbits is
limited to the size of the system. And, second, when the particle
crosses the equatorial line, the vertical reaction of the surface
points downward and the particle may lose contact with the
surface. \newline
For all these reasons, we believe that the cone problem has a
great academic potential and deserves adequate attention in
General Physics courses.

\section{Conclusions}
\label{sec:conclusions}

The understanding of the physical mechanisms of many complicated 
phenomena requires a drastic simplification in a first 
approach. Here we have isolated two basic ingredients, the conical
geometry and the gravitational field, present in 
different natural formations such as volcanoes or tornadoes, and 
in a series of technical devices such as pipes, 
particle precipitators or centrifugal separators. 

We have studied the evolution of a particle sliding on the 
inside of a conical surface, with opening angle $\phi_0$,
under the action of gravity.  
This conservative system has two constants of motion,
the total energy, $\cal H$, and the vertical component 
of the angular momentum, $L_z$, which makes it 
integrable. The initial values of both constants 
determine the trajectory described in its
four-dimensional phase space, which in general is quasiperiodic
and it fills up a 2-torus.

Every orbit is closed when it is projected
in the radial plane $(r,\dot{r})$, running between
the two extreme values, $r_{min}$ and $r_{max}$.
For a quasiperiodic trajectory, the oscillations in 
the angular part are not synchronized with the radial ones.
It means that wherever the particle is released on the 
inside of the cone, it will come back as closely
as desired to the same point. Only when the relation 
between the frequencies of these two (radial and angular)
uncoupled modes is a rational number, $p/q$, we will have
a periodic motion. This is satisfied when 
$\frac{p}{q}=k(E)\sin\phi_0$, where $k(E)$ is a 
parameter dependent only on the product 
$E\sim{\cal H} L_z^{-2/3}$ (see eqs. (\ref{eq:univ-rel})
and  (\ref{eq:radial-e})) and given by the expression
(\ref{eq:k(E)}). The range of $k(E)$ has been
numerically calculated in Fig. 6 and it varies in the interval
$\sqrt{3}<k(E)<2$, where the value $k=\sqrt{3}$ is reached for
small oscillations about the circular motion
and $k=2$ when the energy of the particle tends to infinity. 

Several universal relations characterizing the dynamics
of the system have been obtained. First, the non dimensional 
equations have been derived by applying different length and time
rescaling transformations. Independently of the initial 
conditions, all the information  about the motion 
is contained in these {\it universal equations}.
Second, the relation between the radial and angular
frequencies has been shown to depend only on the parameter $k(E)$.
It has been reinterpreted as an {\it universal dynamical relation}
of the number of revolutions that the particle performs around the cone for
each radial oscillation. Third, every trajectory develops 
on a truncated part of the cone. The characteristic lengths 
and extreme velocities defining the orbit satisfy a {\it universal
geometrical relation}. Finally, some relations of the transfer of energy
between the orbital, radial and potential energy have been presented.

The application of these relations to some realistic systems can
be used to test if the corresponding underlying 
dynamics is understandable as caused by the 
combined action of the gravitational field and a
force perpendicular to the motion, which could be identified as 
the reaction force of the conical profile associated with
the phenomenon under study.
It is our aim to explore this direction in a future work. 

\newpage
\appendix
\section{Appendix: Periodic Trajectories of the Motion}
\label{appendix1} Periodic motions of a dynamical system are
represented by closed orbits. They play a fundamental role in
understanding the topology and orbit organization of the phase space,
but there are no general methods to find them. Here we sketch a
procedure that simplifies the localization of periodic orbits in the
present system. It consists of an adequate projection on a phase
space of a lower dimension, where the property of periodicity is
conserved. \newline By a {\it conserving periodicity projection}
we mean that if an orbit is periodic in the space of projection
then it is also periodic in the original higher dimensional
system. If so, there exists a bi-univocal relation between the
periodic orbits of both spaces, and it is enough to find them in
the space of projection. Thus the problem is enormously
simplified.

The motion of the particle inside the cone is governed by four
autonomous differential equations of first order, given by
Hamilton's equations of the system:
\begin{eqnarray}
 \dot{r} = \;\;\frac{\partial {\cal H}}{\partial p_r} & = &
 \frac{p_r}{m}, \label{eq:h1}\\
 \dot{\theta} = \;\;\frac{\partial {\cal H}}{\partial p_{\theta}}
 & = & \frac{p_{\phi}}{mr^2\sin^2\phi_0},\\
 \dot{p_r} = -\frac{\partial {\cal H}}{\partial r} & = &
 \frac{-p^2_{\phi}}{mr^3\sin^2\phi_0} - mg\cos\phi_0, \\
 \dot{p_{\theta}} = -\frac{\partial {\cal H}}{\partial \theta} & = &
 0. \label{eq:h2}
\end{eqnarray}
Hence a trajectory of the system evolves in the four-dimensional
phase space with coordinates $(r, \theta, p_r, p_{\theta})$, or,
equivalently, with coordinates $(r, \theta, \dot{r},
\dot{\theta})$ if the change of variables $(p_r,
p_{\theta})\rightarrow (\dot{r}, \dot{\theta})$, given by Eqs.
(\ref{eq:momenta1}-\ref{eq:momenta2}), is performed.\newline As
the dynamical equations fulfil the regular conditions of
continuity and smoothness, the field is uniquely determined on
each point, and the flow has no singularities in the region of
phase space with physical significance. That is, those points
whose coordinates verify: $r>0$, $0\leq\theta<2\pi$ and
$0<\phi_0<\frac{\pi}{2}$. \newline A phase path will be closed,
and then periodic, when it visits the same point twice and
in consequence, an infinite number of different times:
\begin{displaymath}
\hbox{A trajectory is closed}\Longleftrightarrow \exists
T>0\rightarrow (r, \theta, \dot{r}, \dot{\theta})_{t=0} = (r,
\theta, \dot{r}, \dot{\theta})_{t=T}.
\end{displaymath}

If we perform the projection $(r, \theta, \dot{r},
\dot{\theta})\rightarrow (r, \dot{r})$, the integral curves on the
new two-dimensional phase space are solutions of Eq.
(\ref{eq:radial-h}) and they are closed orbits as can be seen,
for instance, in Fig. 3. Therefore, for this particular
projection, all the four-dimensional trajectories are projected on
two-dimensional periodic orbits. This is a non conserving
periodicity projection.

By simple inspection of the problem it is obvious that periodic
orbits of the equations system (\ref{eq:h1}-\ref{eq:h2}) are those
verifying $(r(t),\theta(t))=(r(t+T),\theta(t+T))$. Take the origin
of time, $t=0$, when the trajectory passes through one of the extreme
points, $r_{min}$ or $r_{max}$, of the radial coordinate. At these
turning points, the radial velocity vanishes, $\dot{r}=0$. As the
angular velocity around the cone $\dot{\theta}$ depends only on
the dynamical constant $L_z$ of the motion and on the radial
coordinate $r$ (see Eq. (\ref{eq:angular})), it follows that $(r,
\theta, \dot{r}=0, \dot{\theta})_{t=0} = (r, \theta, \dot{r}=0,
\dot{\theta})_{t=T}$, and the trajectory is also closed in the
total four-dimensional space.\newline Therefore, in order to find
the periodic motions of the whole four-dimensional system we must
study the conserving periodicity projection: $(r, \theta, \dot{r},
\dot{\theta})\rightarrow (r, \theta)$. The periodic orbits are in
this case all those verifying that the ratio between its radial,
$\omega_r$, and, angular, $\omega_{\theta}$, frequencies is a
rational number:
\begin{equation}
\hbox{Periodic motion}\;\; \Leftrightarrow\;\;
\frac{\omega_r}{\omega_{\theta}}=\frac{p}{q}\;\;\hbox{ with }
p,q\in \mathrm{Z}. \label{eq:cond-periodic}
\end{equation}
Note that this conclusion is well-established in literature
\cite{goldstein}, and is a consequence of the integrability of
the Hamiltonian. The present four-dimensional system has two
independent global constants, namely, the energy, $\cal H$, and
the vertical component of the angular momentum, $L_z$, which allow
us to rewrite the Hamiltonian in action-angle coordinates. Hence the
trajectories are $2$-frequency quasi-periodic and, in general,
they fill up a $2$-torus. The angular velocities specifying the
motion on the $2$-torus are $\omega_r$ and $\omega_{\theta}$, and
the condition for having a periodic orbit is given by Eq.
(\ref{eq:cond-periodic}).

\newpage
\begin{center} \bf Figure Captions \end{center}\par

{\bf 1.} Geometrical coordinates $(r,\theta)$ defining the
dynamics of the particle of mass $m$, which slides under the
action of the gravitational field $g$ on the inside of a cone 
with opening angle $\phi_0$.

{\bf 2.} Plot of the effective radial potential energy, $U(r)$,
for $L_z=1$, $\phi_0=\frac{\pi}{4}$ and {m=1}.

{\bf 3.} Pattern of solutions of the radial equation (\ref{eq:radial})
for $\phi_0=\frac{\pi}{4}$.
These are closed curves between $r_{min}$ and $r_{max}$ taking the
maximal radial velocity for $r=r_0$. ({\bf a}) $L_z=0.1$, $r_0=0.142$ ;
({\bf b}) $L_z=1$, $r_0=0.660$ ; ({\bf c}) $L_z=10$, $r_0=3.066$.

{\bf 4.} Small radial oscillations about the circular motion for $L_z=1$:
$r\sim r_0+\Delta r$, $\Delta r_{max}\sim 0.05$. 
({\bf a}) $\omega_r>\omega_0$ for $\phi_0=\frac{\pi}{4}>\bar{\phi}_0$,
$r_0=0.660$. ({\bf b}) $\omega_r<\omega_0$ for 
$\phi_0=\frac{\pi}{6}<\bar{\phi}_0$, $r_0=0.778$. 

{\bf 5.} Large radial oscillations about the circular motion for $L_z=1$:
$r\sim r_0+\Delta r$, $\Delta r_{max}\sim 1.5$. 
({\bf a}) $\omega_r>\omega_{\theta}$ for $\phi_0=\frac{\pi}{4}>\bar{\phi}_0$,
$r_0=0.660$. ({\bf b}) $\omega_r<\omega_{\theta}$ for 
$\phi_0=\frac{\pi}{6}<\bar{\phi}_0$, $r_0=0.778$. 

{\bf 6.} Numerical calculation of 
$k=\frac{\omega_r}{\omega_{\theta}\sin\phi_0}$ as a function of $r\_relation$:
{\bf (+)} $r\_relation = \frac{r_{min}}{r_0}$ and 
{\bf (x)} $r\_relation = \frac{r_0}{r_{max}}$. This curve
is independent of $L_z$ and $\phi_0$. In this case,
the computation has been performed for $L_z=1$ and $\phi_0=\frac{\pi}{4}$.

{\bf 7.} Characteristic lengths of a typical orbit developing on 
the truncated cone defined by $R_{min}$, $R_{max}$ and $H$.


\newpage
\begin{thebibliography}{99}

\bibitem{turcotte} D.L. Turcotte and G. Schubert, {\it Geodynamics:
Applications of Continuum Physics to Geological Problems},
Section 9.6, John Wiley \& Sons (1982).

\bibitem{hancock} {\it The Oxford Companion to the Earth}, Edited by
P.L. Hancock and B.J. Skinner, page 1089, Oxford University Press (2000).

\bibitem{bluestein} H. Bluestein, {\it Tornado Alley: Monster
Storms of the Great Plains}, Oxford University Press (1999).

\bibitem{lalor} C. B. Lalor and A. J. Hickey, 
"Pharmaceutical aerosols for delivery 
of drugs to the lungs", in {\it Physical and Chemical Properties of Aerosols}, 
pag.391, Edited by I. Colbec, Blackie Academic \& Professional (1998).

\bibitem{masters} G. M. Masters, {\it Introduction to Environmental 
Engineering and Science}, 2nd edit. Prentice Hall Inc. (1998).

\bibitem{xiang} R. Xiang, S.H. Park and K.W. Lee, "Effects of 
cone dimension on cyclone performance", Aerosol Science {\bf 32}, 
549-561 (2001).

\bibitem{alhawaj} O. Al-Hawaj, "A numerical study of the 
hydrodynamics of a falling liquid film on the internal surface of
a downward tapered cone", Chemical Engineering Journal {\bf 75},
177-182 (1999).

\bibitem{fluids} G.K. Batchelor, "An Introduction to
Fluid Dynamics", section 7.8, Cambridge University Press (1994).

\bibitem{bak} P. Bak, C. Tang, and K. Wiesenfeld,
"Self-organized criticality: an explanation of the 1/f noise",
Phys. Rev. Lett. {\bf 59}, 381–384 (1987).

\bibitem{kibble} T.W.B. Kibble, {\it Classical Mechanics}, page 263,
Longman Scientific \& Technical (1985).

\bibitem{andronov} A.A. Andronov, A.A. Vitt and S.E. Khaikhin,
{\it Theory of Oscillators}, Dover Publications (1996).

\bibitem{goldstein} H. Goldstein, {\it Classical Mechanics},
Addison-Wesley, Reading (1980).

\end{thebibliography}
\end{document}